\def\lj{{\lambda_J}}
\def\bSG{{\beta_{_{c}}}}

\def\Pt{{P_{mean}(t)}}
\def\fML{\varphi_{\scriptscriptstyle Max}^{\scriptscriptstyle L}}
\def\fMR{\varphi_{\scriptscriptstyle Max}^{\scriptscriptstyle R}}

\def\V{{\zeta}}
\def\iT{{i_{\scriptscriptstyle{T}} }}
\def\gT{{\gamma_{\scriptscriptstyle{T}} }}

\def\Pt{\overline{P}(t) }

\documentclass[12pt]{iopart}

\expandafter\let\csname equation*\endcsname\relax
\expandafter\let\csname endequation*\endcsname\relax

\usepackage{color}
\usepackage{graphicx}
\usepackage{epstopdf}
\usepackage{amsmath,amssymb,wasysym}
\usepackage[sort&compress,numbers]{natbib}
\usepackage{mathrsfs}

\begin{document}
\date{\today}

\title[Effects of L\'evy noise on the dynamics of SG solitons in long JJ]{Effects of L\'evy noise on the dynamics of sine-Gordon solitons in long Josephson junctions}

\author{Claudio Guarcello$^{1,2}$, Davide Valenti$^1$, Angelo Carollo$^{1,2}$, Bernardo Spagnolo$^{1,2,3}$}

\address{$^1$ Dipartimento di Fisica e Chimica, Group of Interdisciplinary Theoretical Physics, Universit\`a di Palermo and CNISM, Unit\`a di Palermo Viale delle Scienze, Edificio 18, 90128 Palermo, Italy}
\address{$^2$ Radiophysics Department, Lobachevsky State University, 23 Gagarin Avenue, 603950 Nizhny Novgorod, Russia}
\address{$^3$ Istituto Nazionale di Fisica Nucleare, Sezione di Catania, Via S. Sofia 64, I-95123 Catania, Italy}
\ead{claudio.guarcello@unipa.it}

\begin{abstract}
We numerically investigate the generation of solitons in current-biased long Josephson junctions in relation to the superconducting lifetime and the voltage drop across the device. The dynamics of the junction is modelled with a sine-Gordon equation driven by an oscillating field and subject to an external non-Gaussian noise. A wide range of $\alpha$-stable L\'evy distributions is considered as noise source, with varying stability index $\alpha$ and asymmetry parameter $\beta$. In junctions longer than a critical length, the mean switching time (MST) from superconductive to the resistive state assumes a values independent of the device length. Here, we demonstrate that such a value is directly related to the mean density of solitons which move into or from the washboard potential minimum corresponding to the initial superconductive state. 
Moreover, we observe: (i) a connection between the total mean soliton density and the mean potential difference across the junction; (ii) an inverse behavior of the mean voltage in comparison with the MST, with varying the junction length; (iii) evidences of non-monotonic behaviors, such as stochastic resonant activation and noise enhanced stability, of MST versus the driving frequency and noise intensity for different values of $\alpha$ and $\beta$; (iv) finally, these non-monotonic behaviors are found to be related to the mean density of solitons formed along the junction.

\end{abstract}

\pacs{85.25.Cp, 05.10.Gg, 72.70.+m, 74.40.−n}

\vspace{2pc} \noindent{\it Keywords}: Metastable states, Mesoscopic systems (Theory), Large deviations in non-equilibrium systems, Stochastic processes

\submitto{\JSTAT}

\maketitle

\section{Introduction}

The problem of detecting a sinusoidal signal corrupted by a Gaussian noise source has been already, at least in principle, completely solved~\cite{Hel68}. However, when the amount of data to process is huge or the signal to noise ratio is too small, the choice of detection strategies becomes crucial. As an
example, we can remember the all-sky all-frequency search for
gravitational waves emitted by a pulsar~\cite{Jar98}, the search for
continuous monochromatic signal in radio astronomy~\cite{Sal89} and
the detection of terahertz radiation~\cite{Mit03}. The use of bistable systems as non-linear devices for signal detection has been recently proposed~\cite{Inc96, Gam98} as a way to tackle these problems. Nonlinear elements inserted in place of linear matched filters may not enhance the overall detection performance~\cite{Gal98}, however can greatly improve the detection strategy and/or reduce the computational and memory overhead.\\
\indent On account of their peculiar properties, Josephson Junctions (JJ) stand out among other nonlinear elements as well suited candidates for signal detection~\cite{Inc96,Khu09}. Notably, they are very fast elements, which can operate at frequency as high as terahertz~\cite{Ozy07}. Moreover, thermal noise effect on JJs can be greatly reduced, cooling down them quite close to the absolute zero, up the quantum noise limit~\cite{CasLeo96}.
Various proposals for the use of JJs as detectors of weak signals in the presence of noise have been put forward, so far. Some of them make use of superconducting quantum interference devices (SQUID)~\cite{Glu02,Gro04}, while others focus instead on the switching from the metastable superconducting state to the resistive running state of the JJ. In the latter case, various approaches have been exploited. The statistical analysis of the switching can be used to reveal weak periodic signals embedded in a noisy environment ~\cite{Fil10,Add12,Add13,Pie15,Add15}. The rate of switching, on the other hand, can provide information about the noise present in an input signal~\cite{Gra08,Urb09,Ank07,Suk07,Kop13}. 
Proposals to use the statistics of the escape times for signal detection have also been put forward~\cite{Gra08,Urb09,Fil10,Add12,Ank07,Suk07,Kop13}. \\
\indent Experimentally, many systems exibit non-Gaussian noise signals~\cite{Hua07,Pel07,Mon84,Shl95,Dyb04,Sou08}. For example, out-of-equilibrium heat reservoir can be regarded as a source of non-Gaussian noise~\cite{Mon84,Shl95,Dyb04}. An example, which can be well modeled by an $\alpha$-stable distribution, can be found in a wireless ad hoc network with a Poisson field of co-channel users~\cite{Sou08}. 
In current base JJ, coupled with non-equilibrium current fluctuations, the effects of non-Gaussian noise on the average escape time from the superconducting metastable state have been experimentally investigated~\cite{Hua07,Pel07}. Moreover, the role of Gaussian~\cite{Gor06,Gor08,Pan04,Aug08,GorPanSpa08,Aug09,Fed07,Fed08,Fed09,Gua15} and non-Gaussian~\cite{Gra08,Urb09,Ank07,Suk07,Aug10,Gua13,Val14,Spa15,GuaVal15} noise sources on both long and short JJs have been theoretically analyzed.\\
\indent In this paper, we study the escape time from a metastable state of a long JJ driven by an external oscillating force and subject to a noise signal.
The main quantity of interest is the \emph{mean switching time} (MST), i.e. the average time the junction needs to switch from the superconducting state to the resistive regime, calculated on a sufficiently large number of numerical realizations. The analysis is performed varying the junction length, the frequency of the driving current, and the amplitude of the noise signal. The noise is modeled by an $\alpha$-stable \textit{L\'evy} distribution. 
These statistics aims at describing real situations~\cite{Sza01} in which the variables show abrupt jumps and very rapid variations, called \emph{L\'evy flights}.\\
\indent L\'evy-type statistics was observed in various research fields,
characterized by the presence of scale-invariance~\cite{Che06,Met00,Uch03,Dub09,Dub05FaNL}. Results on
L\'evy flights were recently reviewed in Ref.~\cite{Dub08}. Moreover, L\'evy statistics allowed to well reproduce several observed evolutions in different scientific areas~\cite{Woy01,Dub07,Dub13}, ranging from zoology~\cite{Sim08,Rey08}, biology~\cite{Wes94,Lan13,Lis15}, population dynamics~\cite{Dub08EPJB,LaC10PRE,LaC10EPJB}, to atmospheric~\cite{Dit99} and geological data~\cite{Shl71}, financial markets~\cite{Man95}, network~\cite{Sim11}, social systems~\cite{Bro06}, signal detection~\cite{Sub15} and solid state physics~\cite{Lur12,Sem12,Bri14,Sub14,Ver14}. An extensive bibliography on $\alpha$-stable distributions/processes and its applications is maintained by Nolan~\cite{Nol15}.\\
\indent The dynamics of phase difference across a long JJ is well described by the sine-Gordon equation, which admits very peculiar wave packet solutions, called \emph{solitons}~\cite{Ust98,But81}. They can be pictured as a $2\pi$ twist of the order parameter $\varphi$, developing and propagating along the junction. These twists carry magnetic flux quanta, i.e. \emph{fluxons}~\cite{McL78,Due82}, which can be observed during the switching towards the resistive state.\\
\indent The paper is organized as follows. In the next section the
sine-Gordon model is presented. In Sec. III the statistical
properties of the L\'evy noise are briefly reviewed, showing some
peculiarities of different $\alpha$-stable distributions. Section IV
gives computational details. In Sec. V theoretical results as a
function of the junction length, the driving frequency, and the
noise intensity are shown and analyzed. Finally, in Sec. VII
conclusions are drawn.

\section{The Model}

\indent The behavior of a long overlap JJ is described by a nonlinear partial differential equation for the order parameter $\varphi$, the
\emph{sine-Gordon} (SG) equation~\cite{Bar82,Lik86}
\begin{equation}
\frac{\hbar}{2e}C\frac{\partial^2 \varphi }{\partial \textsc{t}^2}+\frac{\hbar}{2e}\frac{1}{R_N}\frac{\partial \varphi }{\partial \textsc{t}}+j_c\sin\varphi =\frac{\hbar c^2}{8\pi ed}\frac{\partial^2 \varphi }{\partial \textsc{x}^2}+\frac{2e}{\hbar}I.
\label{SGnonNorm}
\end{equation}
\indent Here $\varphi$ is the phase difference between the wave functions describing the superconducting condensate in the two electrodes. Hereafter, small (capitol) letters are used to indicate normalized (non-normalized) coordinates and current terms (with respect to the critical value $I_c$). In our JJ model, the junction is extended along only one direction, defined as $\textsc{x}$, and can be considered ``short'' along the other directions, so that a magnetic field through the junction points orthogonally to the $\textsc{x}$ direction.
In Eq.~(\ref{SGnonNorm}) $j_c$ is the critical current density, $\hbar$ is the Planck constant, $e$ is the value of the electron charge, $c$ is the speed of light, $R_N$ and $C$ are the effective normal resistance and capacitance of the junction, respectively. The magnetic penetration $d=\lambda_L+\lambda_R+t_i$ is the sum of the London depths in the left and right superconductors $\lambda_L$ and $\lambda_R$, respectively, and the interlayer thickness $t_i$. The current $I$ includes both the bias and the fluctuating current.\\
\indent The SG equation can be recast~\cite{Bar82} in terms of the dimensionless $x=\textsc{x}/\lj$ and $t=\textsc{t}\omega_J$ variables, that are the space and time coordinates normalized to the Josephson penetration depth $\lj=\sqrt{\hbar c^2/\left ( 8\pi edj_c \right )}$ and to the inverse of the characteristic frequency $\omega_J=\left ( 2e/\hbar \right )I_cR_N$ of the junction, respectively. Using dimensionless variables, the SG reads
\begin{eqnarray}
\bSG \varphi_{tt}(x,t)+\varphi_{t}(x,t)-\varphi_{xx}(x,t) = i_b(x,t)-\sin(\varphi(x,t))+i_{f}(x,t).
\label{SG}
\end{eqnarray}
\indent Here a simplified notation has been used, with the subscript of $\varphi$ indicating the partial derivative in that variable. This notation will be
used throughout all paper. In Eq.~(\ref{SG}), the fluctuating current density $i_{f}(x,t)$ is the sum of two contributions,
a Gaussian thermal noise $\iT(x,t)$ and an external non-Gaussian noise source $i_{nG}(x,t)$
\begin{equation}
i_{f}(x,t) = \iT(x,t) + i_{nG}(x,t) .
\label{fluct.current}
\end{equation}
The coefficient
\begin{equation}
\bSG=\omega_J R_NC
\label{fluct.current}
\end{equation}
is the Stewart-McCumber parameter. The junctions with $\bSG\ll 1$ have small capacitance and/or small resistance, and are highly damped (overdamped JJ). In contrast, the junctions with $\bSG\gg 1$ have large capacitance and/or large resistance, and are weakly damped (underdamped JJ).
The terms $i_b(x,t)$ and $\sin(\varphi)$ of Eq.~(\ref{SG}) are, respectively, the bias current and supercurrent, both
normalized to the JJ critical current $I_c$. Eq.~(\ref{SG}) is solved imposing the following boundary conditions
\begin{equation}
\varphi_{x}(0,t) = \varphi_{x}(L,t) = \Gamma,
\label{boundary}
\end{equation}
where $\Gamma$ is the normalized external magnetic field. Hereinafter we impose $\Gamma = 0$.\\
\indent The two-dimensional time-dependent tilted potential, named \emph{washboard potential}, is given by
\begin{equation}
U ( \varphi ,x,t )=1-\cos(\varphi)-i_b( x,t ) \thinspace \varphi,
\label{Washboard}
\end{equation}
\begin{figure*}[t!!]
\centering
\includegraphics[width=0.50\textwidth]{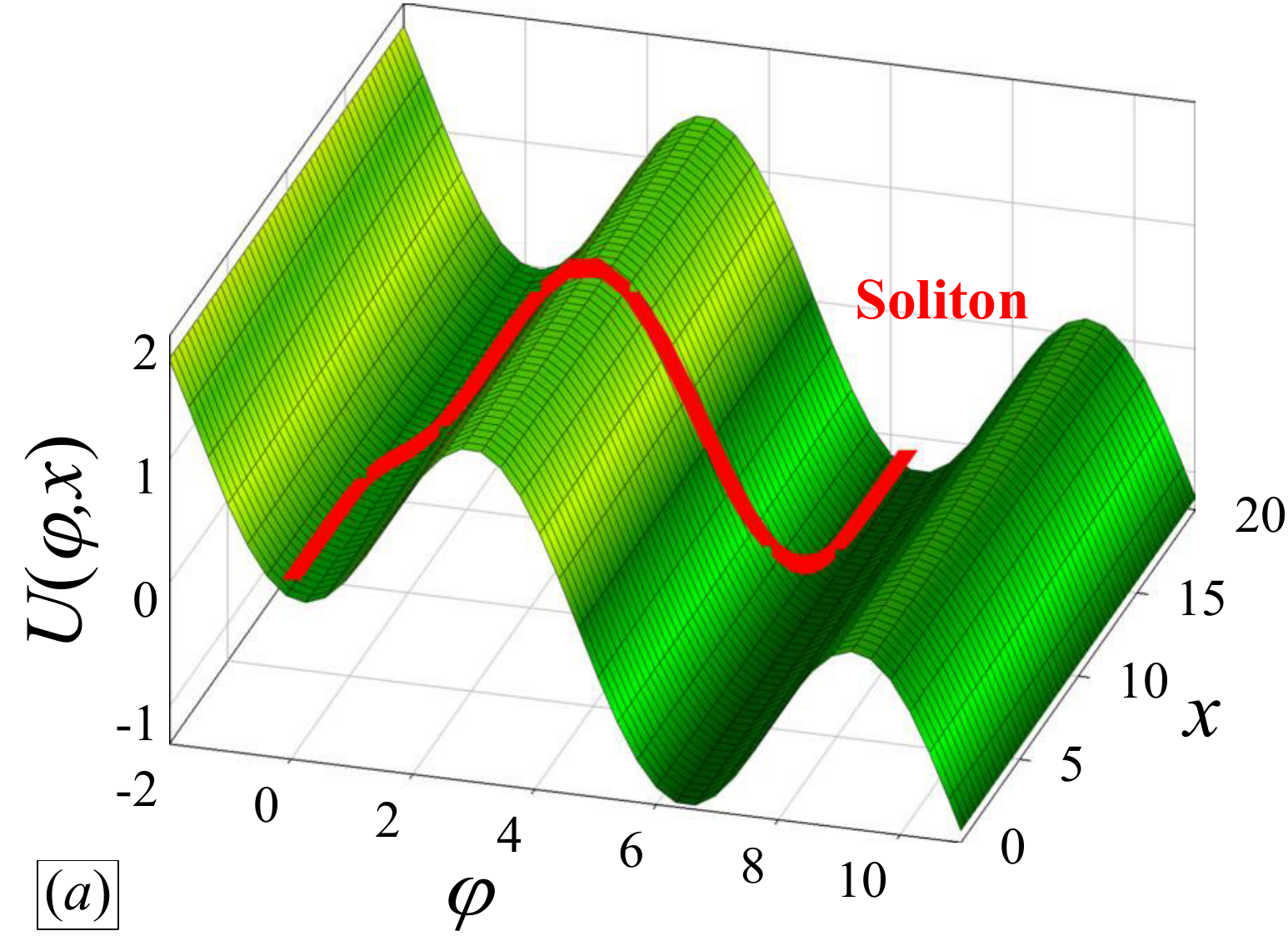}
\includegraphics[width=0.49\textwidth]{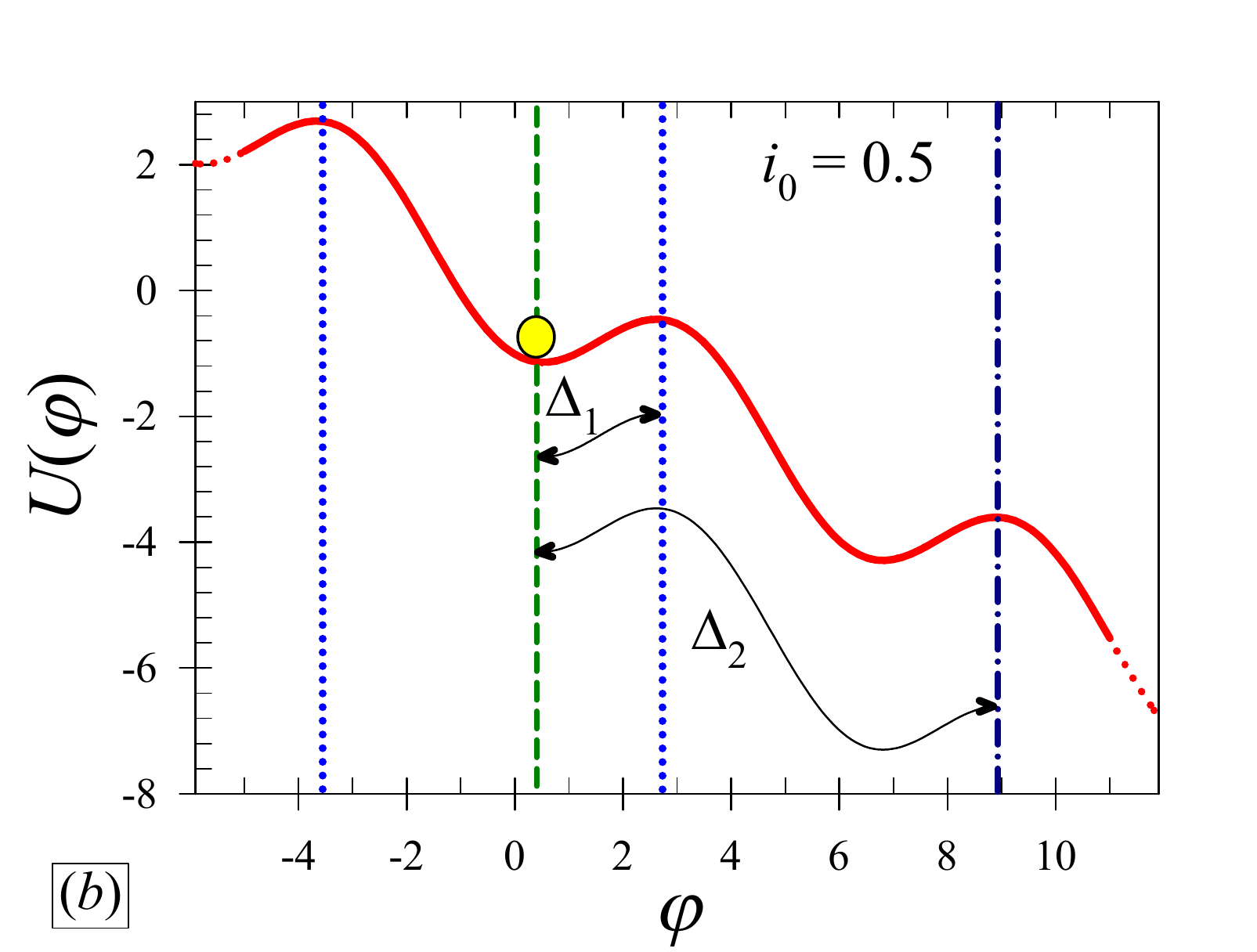}
\caption{(Color online) a) Washboard potential in a fixed istant of its dynamics with a soliton located between two adjacent valleys. b) Cross section of a washboard potential with slope $i_0=0.5$. It is also shown the initial position (bottom of the potential well) of a cell. Blue dotted lines indicate the left and right thresholds. The picture shows also the distances, along the potential profile, $\Delta_{1}$ and $\Delta_{2}$ between the initial minimum and, respectively, the two successive maxima on the right.}
\label{Fig1}
\end{figure*}
and shown in panel \emph{a} of Fig.~\ref{Fig1}. In the same figure is shown a phase string along the potential profile given in Eq.~(\ref{Washboard}).
Specifically, the washboard potential is composed by a periodical sequence of
peaks and valleys, with minima and maxima satisfying the following conditions
\begin{eqnarray}
\label{min_max}
\varphi_{min}=\arcsin( i_b(x,t))+2n\pi \qquad \varphi_{max}=(\pi-\arcsin( i_b(x,t)))+2n\pi
\end{eqnarray}
with $n=0,\pm1,\pm2,\ldots$.\\
\indent The bias current is given by
\begin{equation}
i_b(x,t) = i_0 + i_{ac} \thinspace \sin(\omega t),
 \label{DrivingCurrent}
\end{equation}
where $i_{ac}$ and $\omega$ are amplitude and frequency (normalized to $\omega_J$) of the dimensionless driving current.
The term $i_0$ is a dimensionless current that, in the phase string picture, represents the initial slope of the potential profile. Increasing the slope of the washboard the height of the right potential barrier reduces. Specifically the expression for the right potential barrier height $\Delta U$ is
\begin{equation}
\Delta U = 2 \left [ \sqrt{1-i_b^2}- i_b\cos^{-1} (i_b)\right ].
 \label{DeltaU}
\end{equation}
\indent The unperturbed SG equation, in the absence of damping, bias, and noise, is given by
\begin{equation}
\varphi_{xx}(x,t)-\varphi_{tt}(x,t)=\sin(\varphi(x,t)).
\label{SGunperturbed}
\end{equation}
This equation admits solutions in the traveling wave form $f=\varphi(x-ut)$~\cite{Bar82}
\begin{equation}
\varphi(x-ut)=4\arctan \left \{ \exp \left [ \pm \frac{\left(x-ut \right )}{\sqrt{1-u^2}} \right ] \right \},
\label{SGkink}
\end{equation}
where $u$ is the propagation velocity normalized to the \emph{Swihart velocity} $\overline{c}$. The Swihart velocity is the characteristic speed of propagation of electromagnetic waves along the junction and is usually more than one order of magnitude smaller than the velocity of light in vacuum.
Eq.~(\ref{SGkink}) represents a single \emph{kink}, or \emph{soliton}, that is a $2\pi$ variation in the phase values. A soliton can be depicted as a string located in two neighboring minima crossing the intermediate maximum once (see red solid line in panel \emph{a} of Fig.~\ref{Fig1}).\\
\indent The signs $+$ and $-$ in Eq.~(\ref{SGkink}) indicate a $2\pi$-kink (soliton) and a $2\pi$-antikink (antisoliton), respectively. In this
framework, according to the equation~\cite{Bar82}
\begin{equation}
\varphi_x=\frac{2e}{\hbar c}d H,
\label{Phi_H}
\end{equation}
$\varphi$ gives a normalized measure of the magnetic flux through the junction, so that Eq.~(\ref{SGunperturbed}) can also
represent the motion of a single \emph{fluxon} (or antifluxon) $\Phi_0=h/2e$.
If the phase evolution shows a single $2\pi$-kink, a single fluxon will propagate along the junction, as shown in Fig.~\ref{Fig1}\emph{a}.\\
\indent The normalized thermal current $\iT\left ({x,t}\right )$ is characterized by the well-known statistical properties of a Gaussian random process
\begin{equation}
\left \langle \iT \left ({x,t}\right ) \right \rangle = 0, \qquad \qquad \qquad \left \langle \iT\left ({x,t}\right )\iT\left ({x',t'}\right ) \right \rangle = 2\gT \delta \left (x-x' \right )\delta \left (t-t' \right ),
\label{WNProperties}
\end{equation}
where $\delta$ is the Dirac delta function and
\begin{equation}\label{WNAmp}
\gT=\frac{kT}{R_N}\frac{\omega_{c}}{I^2_c}=\frac{2e}{\hbar}\frac{kT}{I_c}=\frac{kT}{E_J}.
\end{equation}
\indent It is worth noting that the noise intensity $\gT$ can be also expressed as the ratio between the thermal energy and the \emph{Josephson coupling energy} $E_J$~(see Eq.~(\ref{WNAmp})).
This equation can be rewritten as
\begin{equation}\label{WNAmp}
\gT=\frac{I_{\scriptscriptstyle{Th}}}{I_c} \qquad ~~ \textrm{where} ~~ \qquad I_{\scriptscriptstyle{Th}}=\frac{2e}{\hbar}kT.
\end{equation}
\indent Here, $ I_{\scriptscriptstyle{Th}}$ is the \emph{equivalent thermal noise current}. Inserting the numerical values, we see that $ I_{\scriptscriptstyle{Th}}\simeq 0.15\mu A$ at liquid helium temperature ($T$=4.2 K).\\
\indent To analyze in more detail the effect of the non-Gaussian noise on the phase dynamics, we will fix the white thermal noise current at a very low intensity. Specifically, the non-Gaussian noise current $i_{nG}(x,t)$ is described by $\alpha$-stable L\'evy distributions.

\section{The L\'evy Statistics}
\label{Levy_noise}
\begin{figure}[t!!]
\centering
\includegraphics[height=5.7cm]{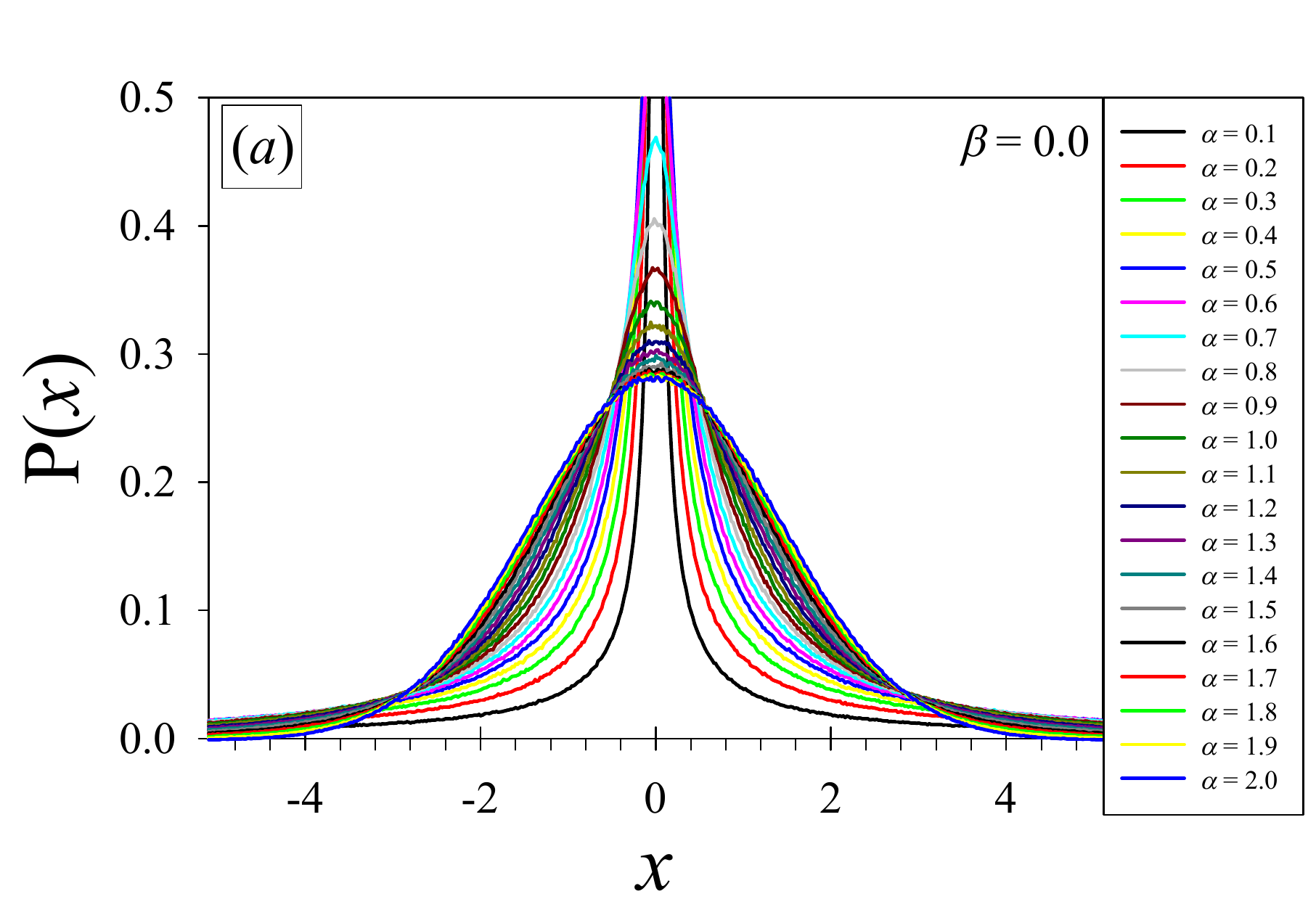}
\includegraphics[height=5.7cm]{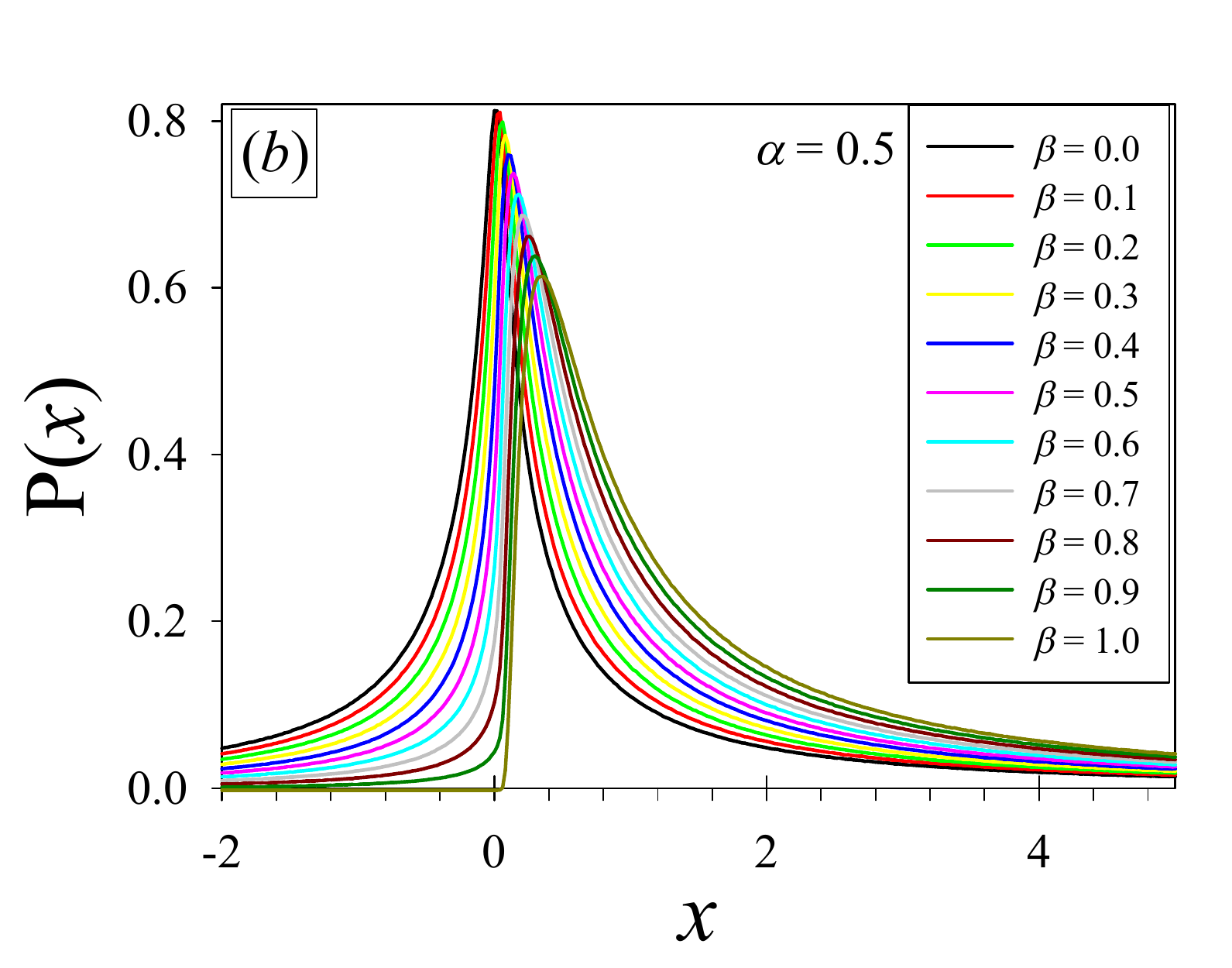}
\caption{(Color online) a) Symmetric stable densities of $S_{\alpha}(1, 0, 0)$ and $\alpha\in(0,2]$. b) Skewed stable densities of $S_{0.5}(1, \beta\ge0, 0)$ and $\beta\in[0,1]$.}
\label{Fig2}
\end{figure}

\vskip-0.2cm
\indent Here we briefly review the concept of $\alpha$-stable L\'evy distributions\cite{Ber96,Sat99,Gne54,deF75,Khi36,Khi38,Fel71}. A random non-degenerate variable $X$ is stable if
\begin{equation}
\forall n\in\mathbb{N}, \exists (a_n,b_n) \in \mathbb{R}^+\times\mathbb{R}: \qquad X+b_n=a_n\sum_{j=1}^{n} X_j, \label{AlfaStable}
\end{equation}
where the $X_j$ terms are independent copies of $X$. Moreover $X$ is strictly stable if and only if $b_n=0 \,\,\, \forall n$. The well
known Gaussian distribution stays in this class. This definition does not provide a parametric handling form of the stable
distributions. The characteristic function, however, allows to deals with them. The general definition of characteristic function for a
random variable $X$ with an associated distribution function $F(x)$ is
\begin{equation}
\phi (u) = \left < e^{iuX} \right > = \int_{-\infty}^{+\infty}e^{iuX}dF(x).
\label{GenerealCharFunc}
\end{equation}
Following this statement, a random variable $X$ is said stable if and only if
\begin{equation}
\exists (\alpha ,\sigma, \beta, \mu )\in\,\,(0, 2]\times \mathbb{R}^+\times [-1, 1]\times \mathbb{R}: \qquad X\overset{d}{=}\sigma Z+\mu,
\label{XStableFunc}
\end{equation}
where $Z$ is a random variable with characteristic function
\begin{eqnarray}
\phi(u)=\left\{\begin{matrix}
\exp \left \{-\left | u \right |^\alpha\left [ 1-i\beta \tan\frac{\pi\alpha}{2}(\textup{sign}u) \right ]\right \} \,\,\,\, \alpha\neq 1\\
\exp \left \{-\left | u \right |\left [ 1+i\beta\frac{2}{\pi}(\textup{sign}u)\log \left | u \right |\right ]\right \} \,\,\,\, \alpha=1
\end{matrix}\right. \label{XCharFunc}
\end{eqnarray}
%
%
%
%
in which
\begin{eqnarray}
\textup{sign}u=\left\{\begin{matrix}
\pm1\,\,\, &u&\gtrless0\\
0 &u&=0
\end{matrix}\right. \label{signU}
\end{eqnarray}
represents the $\textup{sign}$ function. These distributions are symmetric around zero when $\beta=0$ and $\mu=0$. In Eq.~(\ref{XCharFunc}) for the $\alpha=1$ case, $0\cdot \log0$ is always interpreted as $\lim_{x\to0} x\log x=0$, giving rise to $\phi(0)=1$.\\
\indent Definition~(\ref{XStableFunc}) of $X$ requires four parameters: a \emph{stability index} (or characteristic exponent) $\alpha\in(0,2]$,
an \emph{asymmetry parameter} $\beta\in[-1,1]$, a scale parameter $\sigma>0$ and a location parameter
$\mu$. The names of these parameters indicate their physical meaning. In most of recent literature, the notation $S_{\alpha}(\sigma, \beta, \mu)$ is used for the class of stable distributions. The stable distributions obtained setting $\sigma=1$ and $\mu=0$ are called \emph{standard}. The case $\beta=0$ gives a
symmetric distribution, while $\alpha$ determines how the tails of the distribution go to zero. The stability index yields the asymptotic long-tail power law for the $x$-distribution, which for $\alpha<2$ is of the $\left | x \right |^{-\left ( 1+\alpha \right )}$ type, while $\alpha=2$ and $\beta=0$ gives a Gaussian distribution.\\
\indent Fig.~\ref{Fig2}\textit{a} shows the bell-shaped probability density functions for the symmetric stable distributions $S_{\alpha}(1, 0, 0)$ with $\alpha\in(0,2]$. As $\alpha$ decreases, the peaks get higher, the regions around the peak become narrower (the so called \emph{limited space displacement}~\cite{Aug10,Val14}), and the tails get heavier. \\
\indent Fig.~\ref{Fig2}\textit{b} shows the probability density functions for the stable distributions $S_{0.5}(1, \beta, 0)$ for $\beta\ge0$. These skewed distributions are characterized by the right tails heavier than the left ones. When $\beta=1$, we say that the distribution is \emph{totally skewed to the right}. The behavior of the $\beta<0$ cases are reflections of the $\beta>0$ ones, with the left tail being heavier. When $\beta=-1$ the distribution is \emph{totally skewed to the left}.\\
\indent We note that in Eq.~(\ref{XStableFunc}), $\tan \left ( \scriptstyle{\frac{\pi \alpha }{2}} \right )=0$ for $\alpha=2$ , so that the characteristic function is real and the distribution is always symmetric, apart from the value of $\beta$. As $\alpha$ decreases the effect of $\beta$ becomes more pronounced, and the left tail gets lighter and lighter for $\beta\to1$. All stable distributions are unimodal, but there is no known formula for the location of the mode. \\
\indent The heavy tails cause the occurrence of events
with large values of $x$, whose probability densities are not negligible. The use of heavy-tailed statistics allows to
consider rare events, corresponding to large values of $x$, because of the fat tails of these distributions. These events
correspond to the L\'evy flights previously discussed. The algorithm used in this work to simulate L\'evy
noise sources is that proposed by Weron~\cite{Wer96} for the implementation of the Chambers method~\cite{Cha76}.

\section{Computational Details}

\label{CompDetail}
\vskip-0.2cm

\indent We study the phase dynamics of a long JJ within the SG overdamped regime, setting $\bSG=0.01$. The time and spatial steps are $\Delta t=\Delta x=0.05$. Due to the stochastic nature of the dynamics of the system, we calculate the mean values of the variables analyzed performing a suitable number $N=(5\cdot10^3-10^5)$ of
numerical realizations (experiments). Throughout the whole paper we use the words \emph{string}, referring to the entire junction, and
\emph{cell} to indicate each of the elements with dimension $\Delta x$ forming the junction. The string at rest within the washboard
potential valley labeled with $n=0$ (see Eq.~(\ref{min_max}) and panel \emph{b} of Fig.~\ref{Fig1}) is chosen as initial condition for solving Eq.~(\ref{SG}), i.e.
$\varphi_0=\arcsin(i_b(0))=\arcsin( i_0)$. We calculate the mean switching time (MST) towards the resistive state, starting from the metastable state (bottom of a potential minimum) corresponding to the superconducting regime. The MST $\tau$ is a nonlinear relaxation time (NLRT)~\cite{Dub04} and represents the mean value of the permanence times of the phase $\varphi$ within the first valley, that is $\varphi\in[\fML\;,\;\fMR]$. The thresholds $\fML$ and $\fMR$ are, respectively, the positions of the left and right maxima which surround the minimum chosen as initial condition (see dotted lines in panel \emph{b} of Fig.~\ref{Fig1}). No absorbing barriers are setted, so that during the entire observation time, $t_{max}$, all the temporary trapping events are taken into account to calculate $\tau$.
The probability $P_{ij}$ that $\varphi\in[\fML\;,\;\fMR]$, in the \textit{i}-th realization for the \textit{j}-th cell, is
\begin{eqnarray}
P_{ij}(t) = \left\{\begin{matrix}
 1 \iff \varphi\in[\fML\;,\;\fMR]\\
\\0 \iff \varphi\notin[\fML\;,\;\fMR].
\end{matrix}\right.
 \label{P_t}
\end{eqnarray}
Summing $P_{ij}(t)$ over the total number $N_{c}=L/\Delta x$ ($L$ is the junction length) of cells and over the number $N$ of realizations, the average probability that the entire string is in the superconducting state at time $t$ can be computed as
\begin{equation}
\Pt = \frac{1}{N \thinspace N_{c}}\sum_{i=1}^{N}\sum_{j=1}^{N_{c}}P_{ij}(t).
\label{P_averaged}
\end{equation}
The MST $\tau$ is therefore calculated as
\begin{equation}
\tau = \int_{0}^{t_{max}} \Pt dt.
 \label{tauNLRT}
\end{equation}
The whole procedure is repeated varying the value of $\alpha$ and
$\beta$, and obtaining the behaviors of the MST $\tau$ in the
presence of different sources of L\'{e}vy noise.

\section{Results}

In this section, we investigate the dependence of the MST $\tau$,
the mean potential difference (MPD) $\V$, and the mean soliton
densities, $n$ and $n_{tot}$, on the junction length $L$, the
driving frequency $\omega$, and the noise amplitude $\gamma$ as the
L\'{e}vy parameters $\alpha$ and $\beta$ are varied.

The mean soliton density $n$ is obtained considering only solitons
partially lying in the potential well chosen as initial condition (see Fig.~\ref{Fig1}\emph{a}). Instead, the total mean
soliton density $n_{tot}$ is obtained
considering all solitons formed along the junction. These densities are calculated taking into account both kinks and antikinks.\\
\indent The noise intensity $\gamma$ refers to the non-Gaussian
component of the noisy current, while hereinafter the intensity of
the thermal contribution is set to $\gT=10^{-4}$.\\
\indent The amplitude of the oscillating component of the driving
current is set to $i_{ac}=0.7$, to allow, within a driving period,
values of $i_b(t)$ greater than 1, corresponding to the absence of
metastable states.

\subsection{Results as a function of $L$}

\begin{figure}[t]
\centering
\includegraphics[width=\textwidth]{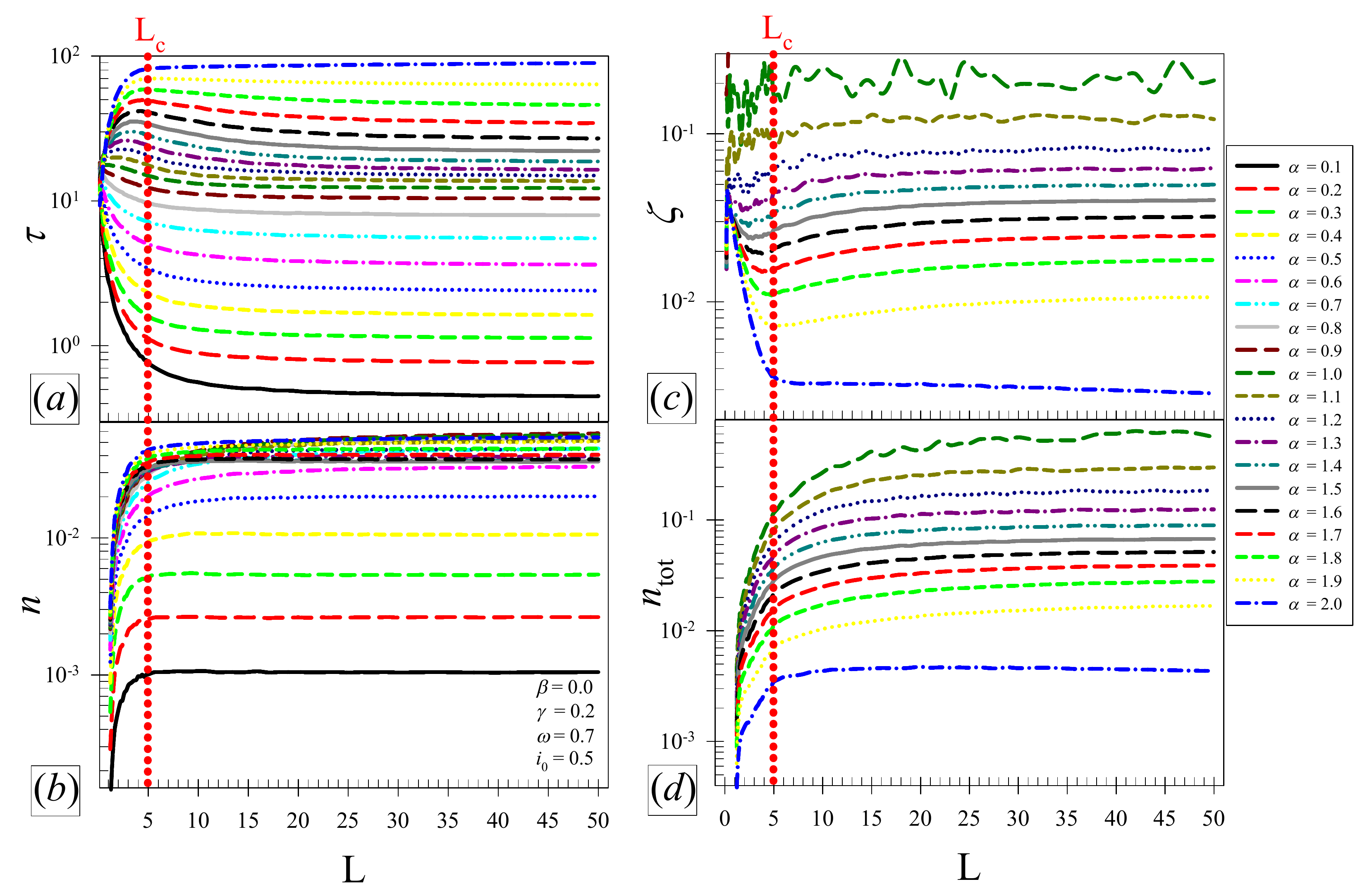}
\caption{(Color online) MST $\tau$ (panel (\textit{a})) and mean soliton density $n$ (panel (\textit{b})) as a function of the junction length $L$ for $S_{\alpha}(1, 0, 0)$ and $\alpha\in(0,2]$. Normalized mean potential difference (MPD) $\zeta$ (see Eq.~(\ref{MeanV})) (panel (\textit{c})) and mean total soliton density $n_{tot}$ (panel (\textit{d})), as a function of the junction length $L$ for $S_{\alpha}(1, 0, 0)$ and $\alpha\in[1,2]$. The values of the other parameters, $\gamma=0.2$, $\omega=0.7$ and $i_0=0.5$, are shown in panel \textit{b} and refer to all panels. The dotted vertical lines mark the nucleation length $L_c\simeq5$. The legend refers to all panels.}
\label{Fig3}
\end{figure}

In this section we study the behavior of a long junction varying its
length $L$. The analysis can be split in two parts: \emph{i}) a
comparison between the MST $\tau$ and the mean soliton density $n$;
\emph{ii}) the study of the \emph{mean potential difference} (MPD)
$\V$ across the junction related to the total mean soliton
density $n_{tot}$. The MPD is in unit
of $\Phi_0$ and is normalized to the junction length $L$.\\
\indent The MSTs as a function of the length $L$, varying the
stability index $\alpha\in(0,2]$, are shown in panel \emph{a} of
Fig.~\ref{Fig3}. The values of the other variables, used to compute
all curves of Fig.~\ref{Fig3}, are $\beta=0$, $\gamma=0.2$,
$\omega=0.7$, and $i_0=0.5$. The values of the MST decrease by
reducing $\alpha$, because the effect of the L\'evy flights become
more relevant reducing $\alpha$. All curves of panel \emph{a} of
Fig.~\ref{Fig3} are characterized by the presence of two different
``dynamical regimes'', in correspondence of values of length below
and above a threshold, namely the \emph{critical}, or
\emph{nucleation}, \emph{length} $L_c$ (dotted red lines in
Fig.~\ref{Fig3}). An initial monotonic behavior is followed by a
constant plateau in the MST curves. Specifically, for $i_0=0.5$,
$L_c\sim5$~\cite{Fed07,Val14}.\\
\indent For values of $L<L_c$, soliton formation is hampered by the
strength of interaction between neighbouring cells. Strings shorter
than the nucleation length are indeed too small/too stiff to form
ripples ample enough to overcome the potential barriers. As a
consequence, below this threshold, the cells forming a string can
only cross a barrier altogether, thereby suppressing chances of
soliton formation. For $L<L_c$, curves in Fig.~\ref{Fig1}\emph{a}
show a monotonic behavior of MSTs as a function of $L$. However,
this dependence changes qualitatively with $\alpha$. MSTs either
increase or decrease depending on whether $\alpha$ is below or above
$1$. When $\alpha\in(1,2]$, moving rigidly a string across a
barrier requires a bigger effort as its length increases. Hence, MST
grows with $L$, up to its maximum in $L\simeq L_c$. For
$\alpha\in(0,1]$ a higher number of cells implies a higher
probability of generating L\'{e}vy flights strong enough to push the
string out the metastable state. Consequently, the MSTs tend to
decrease as $L$ increases.\\
\indent Conversely, long strings move from a potential well by
formation of kinks, antikinks and/or kink-antikink pairs. For
$L>L_c$ a saturation effect is evident~\cite{But81,Cas96,
Sim97,Fed07,Val14}. The MST reaches an almost constant value,
indicating that the dynamics of the switching events is independent
of the JJ length and these events are guided by the solitons. To
explain this behavior, Valenti \emph{et al.}~\cite{Val14} proposed a
subdomains structure of the string, in which each subdomain is
composed by an amount of cells of total size approximately equal to
the critical length~\cite{But81}. The entire string can be thought as the sum of
these subdomains and the overall escape event results to be the
superimposition of the escape events of each single subdomain.
Consequently, increasing the junction length, the total MST is
constant because it is equal to the time evolution of the individual
subdomain. This means that the amount of generated solitons grows
linearly with the amount of subdomain, that is with the length of
the junction. Inspired by this picture, we calculate the \emph{mean
soliton density} $n$. First, the soliton density is calculated
considering, in each experiment, all kinks and anti-kinks partially lying in
the initial metastable state, which is the same well used to
calculate the MSTs. Then, the mean soliton density $n$ is obtained
by averaging over the total number of experiments $N$. According to
Ref.~\cite{Val14}, we observe a saturation effect in the
behavior of $n$ as a function of $L$. These curves are shown in
panel \emph{b} of Fig.~\ref{Fig3} varying $\alpha\in(0,2]$. For
$L<L_c$, $n$ rapidly rises up to reach a plateau for $L>L_c$. The
saturation in the MSTs can be therefore read through the constant
value assumed by the mean soliton density $n$ for long JJ.\\
\indent Depending on the junction parameters, the experimentalists
can choose what to measure, either the voltage or the MST. For
example, measurements of mean ``noise-induced'' voltage may be
performed easily for strong to moderate damping ($\bSG<1$). In fact, with
increasing $\bSG$ (underdamped regime) the maximal value of current $I_{max}<I_c$, above
which the switching to the running state occurs, decreases, and the
measured voltages become also smaller (see Fig.~8 of
Ref.~\cite{Fed07}). Moreover, for large $\bSG$ the measurements of MSTs to the resistive running state may be performed
more easily as the time between sequential single flux quantum pulses. \\
\indent According to the \emph{a.c. Josephson relation}~\cite{Jos62,Jos74}, when the string rolls down along the potential, a non-zero mean voltage across the junction appears
\begin{equation}\label{acJosEffect}
\frac{1}{2\pi}\frac{\mathrm{d} \varphi \left ( x,t \right )}{\mathrm{d} t}=\frac{V\left ( x,t \right )}{\Phi _0}.
\end{equation}
\indent The potential difference across the JJ (normalized to $\Phi_0$), in the \textit{i}-th realization at the time $t$, is
\begin{equation}\label{MeanV_norm}
\frac{V_i\left ( t \right )}{\Phi_0}=\int_{0}^{L}\frac{V_i\left ( x,t \right )}{\Phi_0}dx=\frac{1}{2\pi }\int_{0}^{L}\frac{\mathrm{d} \varphi \left ( x,t \right )}{\mathrm{d} t}dx.
\end{equation}
\indent Therefore, the mean voltage for the \textit{i}-th realization is
\begin{equation}\label{MeanV_i}
V_i=\frac{\left \langle V_i \right \rangle}{\Phi_0}=\frac{1}{t_{max}}\int_{0}^{t_{max}}\frac{V_i\left ( t \right )}{\Phi_0}dt.
\end{equation}
\indent Averaging over the total number of experiments $N$ and dividing for the JJ length $L$, we obtain the MPD $\V$
\begin{equation}\label{MeanV}
\V=\frac{\overline{V}}{L}=\frac{1}{L}\left ( \frac{1}{N}\sum_{i=1}^{N}V_i\right ).
\end{equation}
\begin{figure}[t]
\centering
\includegraphics[width=\textwidth]{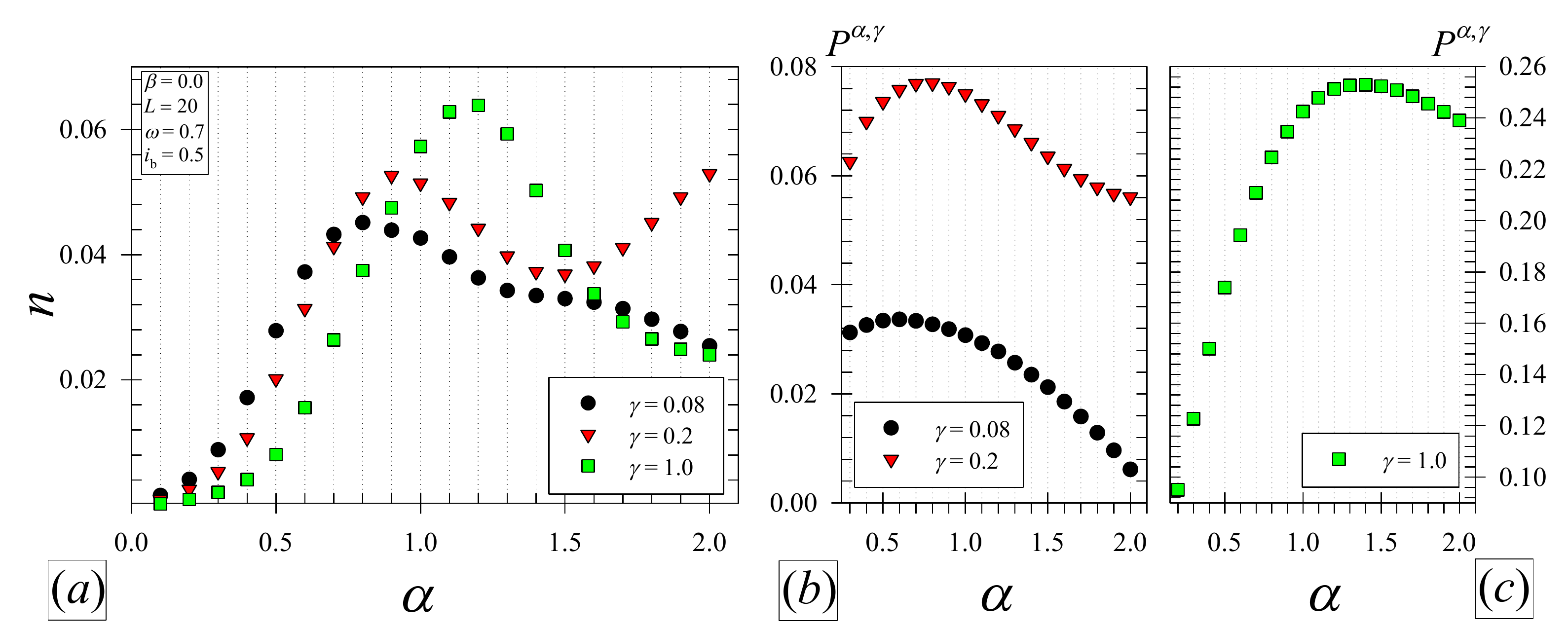}
\caption{(Color online) \textit{a}) Mean soliton density $n$ as a function of the stability index $\alpha$ for $S_{\alpha}(1, 0, 0)$ and $\alpha\in(0,2]$ varying the noise intensity $\gamma=0.08, 0.2, 1.0$. The values of the other parameters are $L=20$, $\omega=0.7$ and $i_0=0.5$. Panels \textit{b}) and \textit{c}) Probability $P^{\alpha,\gamma}$ to obtain values of the random variable $x$ within the range $(\Lambda_1^{\alpha,\gamma},\Lambda_2^{\alpha,\gamma})$ (see Eq.~(\ref{P_LxL})) as a function of $\alpha$, for $\gamma=0.08, 0.2$ (panel \textit{b}) and $\gamma=1.0$ (panel \textit{c}). Here $\Lambda_{1}^{\alpha,\gamma}=\Delta_{1}/D^{\alpha,\gamma}$ and $\Lambda_{2}^{\alpha,\gamma}=\Delta_{2}/D^{\alpha,\gamma}$, where $\Delta_{1}$ and $\Delta_{2}$ are, respectively, the distances between the initial minimum and the two successive maxima on the right (see panel \emph{b} of Fig.~\ref{Fig1}) and $D^{\alpha,\gamma}=(2\gamma)^{1/\alpha}$ is the L\'evy noise amplitude.}
\label{Fig4}
\end{figure}
\indent The curves of the MPD $\V$ as a function of $L$, varying
$\alpha\in[1,2]$, are presented in panel \emph{c} of
Fig.~\ref{Fig3}. The curves for $\alpha<1$ are strongly fluctuating
and are not included here. The results in Fig.~\ref{Fig3}\emph{c}
highlight an ``inverse'' behavior of the MPDs in comparison with the
MSTs~\cite{Fed07}. Decreasing $\alpha$, the time derivative $\varphi_t$ grows because of the L\'evy flights and the MPD values increase (see Eq.~(\ref{acJosEffect})).
Moreover, every MPD curve decreases for small lengths ($L\lesssim
L_c$) up to a minimum above which it tends to a roughly constant
value. In fact, for $L\lesssim L_c$ as the length of the junction increases, the stiffness of the string grows and then the time derivative $\varphi_t$ reduces, giving rise a decreasing of MPD (see Eq.~(\ref{acJosEffect})).\\
\indent The constant value of $\V$ for $L>L_c$ is related to the saturation effect in the values of the mean soliton density. This is calculated by ensamble average of the number of kinks and anti-kinks formed along the string, without
focusing only on the initial metastable state. We call this quantity the \emph{total mean soliton density} $n_{tot}$. 
The values of $n_{tot}$ as a function of $L$, varying
$\alpha\in[1,2]$, are shown in panel \emph{d} of Fig.~\ref{Fig3}. We
observe that $n_{tot}$ rapidly grows for $L<L_c$, but above this
threshold length, i.e. $L>L_c$, the $n_{tot}$ value tends to a
constant value. Comparing the curves of panels \emph{c} and \emph{d}
for $L>L_c$, we observe that $n_{tot}$
closely maps the behavior of the MPD $\V$. We note that the saturation effect in both $n$ and $n_{tot}$ is proportional to the number of subdomains giving rise to solitons along the string.\\
\indent The curves in panel \emph{b} of Fig.~\ref{Fig3} hide a
non-monotonic behavior as a function of $\alpha$. The values of $n$
versus $\alpha$ for $L=20$, $\omega=0.7$ and $i_0=0.5$ are shown in
panel \emph{a} of Fig.~\ref{Fig4}, for three different noise
intensities, namely $\gamma=0.08, 0.2, 1.0$. The data for
$\gamma=0.2$ (red triangles) are selected from
Fig.~\ref{Fig3}\emph{b} for $L=20$, and have a maximum for
$\alpha=0.9$. The position of this maximum changes with the noise
amplitude. Specifically, it is centered in $\alpha=0.8$ for
$\gamma=0.08$, and in $\alpha=1.2$ for $\gamma=1.0$. The mean
soliton density $n$ takes into account the solitons formed with
respect to the first washboard valley. We can rightly assume that
these solitons are most likely generated by direct flights from the
first to the second washboard valley. We define the probability of
``jumping'' from the initial minimum to the second one as
\begin{equation}\label{P_LxL}
P^{\alpha,\gamma}=\textup{Prob}\left \{ \Delta_{1}<D^{\alpha,\gamma}x<\Delta_{2} \right \}=\textup{Prob}\left \{ \Lambda^{\alpha,\gamma}_{1}<x<\Lambda^{\alpha,\gamma}_{2} \right \}.\\
\end{equation}
\indent Here $\Delta_{1}$ and $\Delta_{2}$ are, respectively, the distances along the potential profile between the initial minimum and the two successive maxima on the right (see panel \emph{b} of Fig.~\ref{Fig1}), $\Lambda_{1}^{\alpha,\gamma}=\Delta_{1}/D^{\alpha,\gamma}$, $\Lambda_{2}^{\alpha,\gamma}=\Delta_{2}/D^{\alpha,\gamma}$ and $D^{\alpha,\gamma}=(2\gamma)^{1/\alpha}$ is the L\'evy noise amplitude~\cite{Aug10}. In panels \emph{b} and \emph{c} of Fig.~\ref{Fig3}, the values of $P^{\alpha,\gamma}$ as a function of $\alpha$ for $\gamma=0.08, 0.2$ (panel \textit{b}) and $\gamma=1.0$ (panel \textit{c}) are shown. The probabilities $P^{\alpha,\gamma}$ show non-monotonic behaviors as a function of $\alpha$, with maxima in $\alpha=0.6$ for $\gamma=0.08$, $\alpha=0.8$ for $\gamma=0.2$ and $\alpha=1.3$ for $\gamma=1.0$. The presence of these maxima in the jump probability $P^{\alpha,\gamma}$, which is a proxy of the probability to generate solitons, accounts for the non-monotonic behaviors of $n$ as a function of $\alpha$ shown in Fig.~\ref{Fig4}\emph{a}.

\subsection{Results as a function of $\omega$}

\begin{figure}[t!!]
\centering
\includegraphics[width=0.49\textwidth]{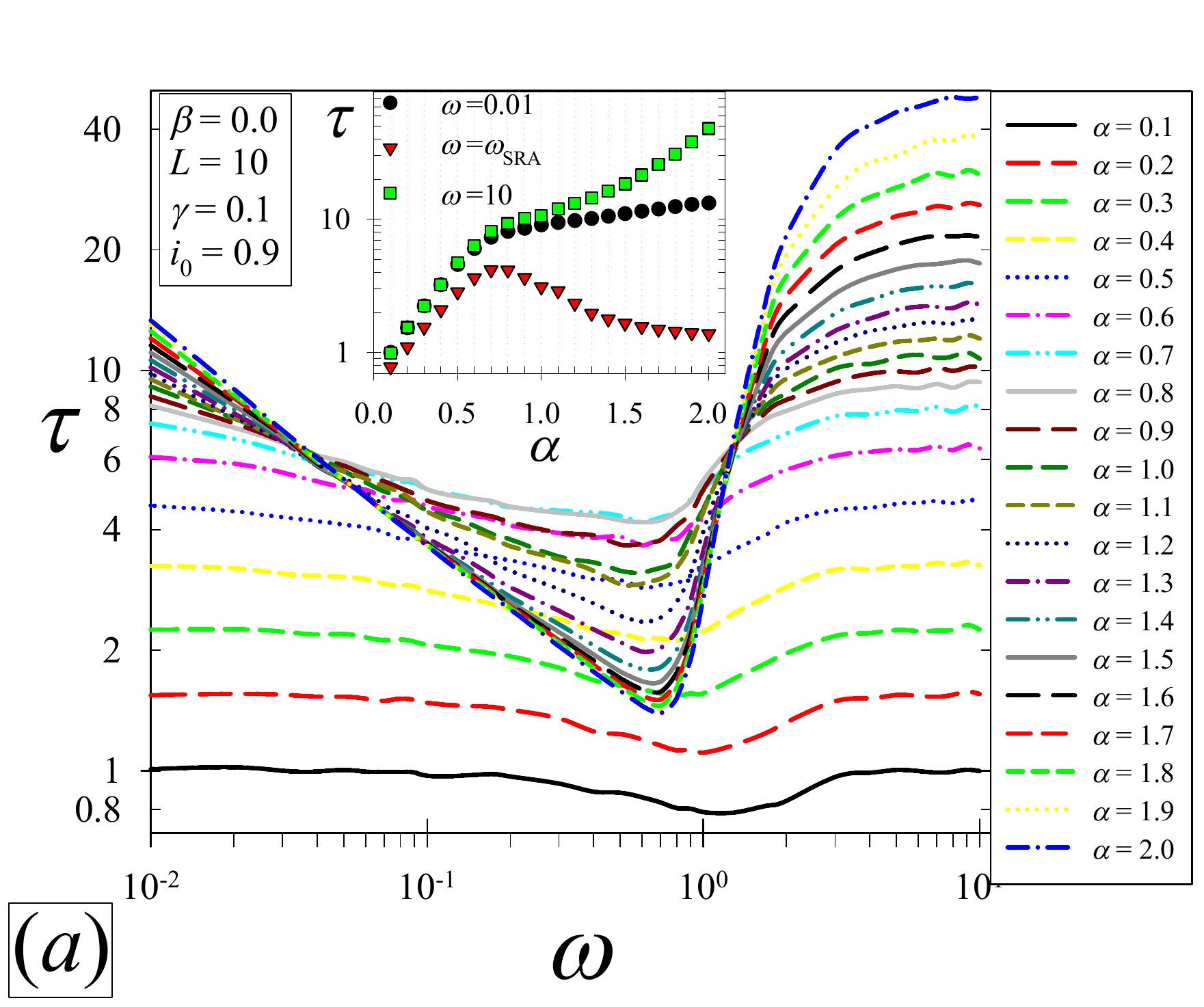}
\includegraphics[width=0.50\textwidth]{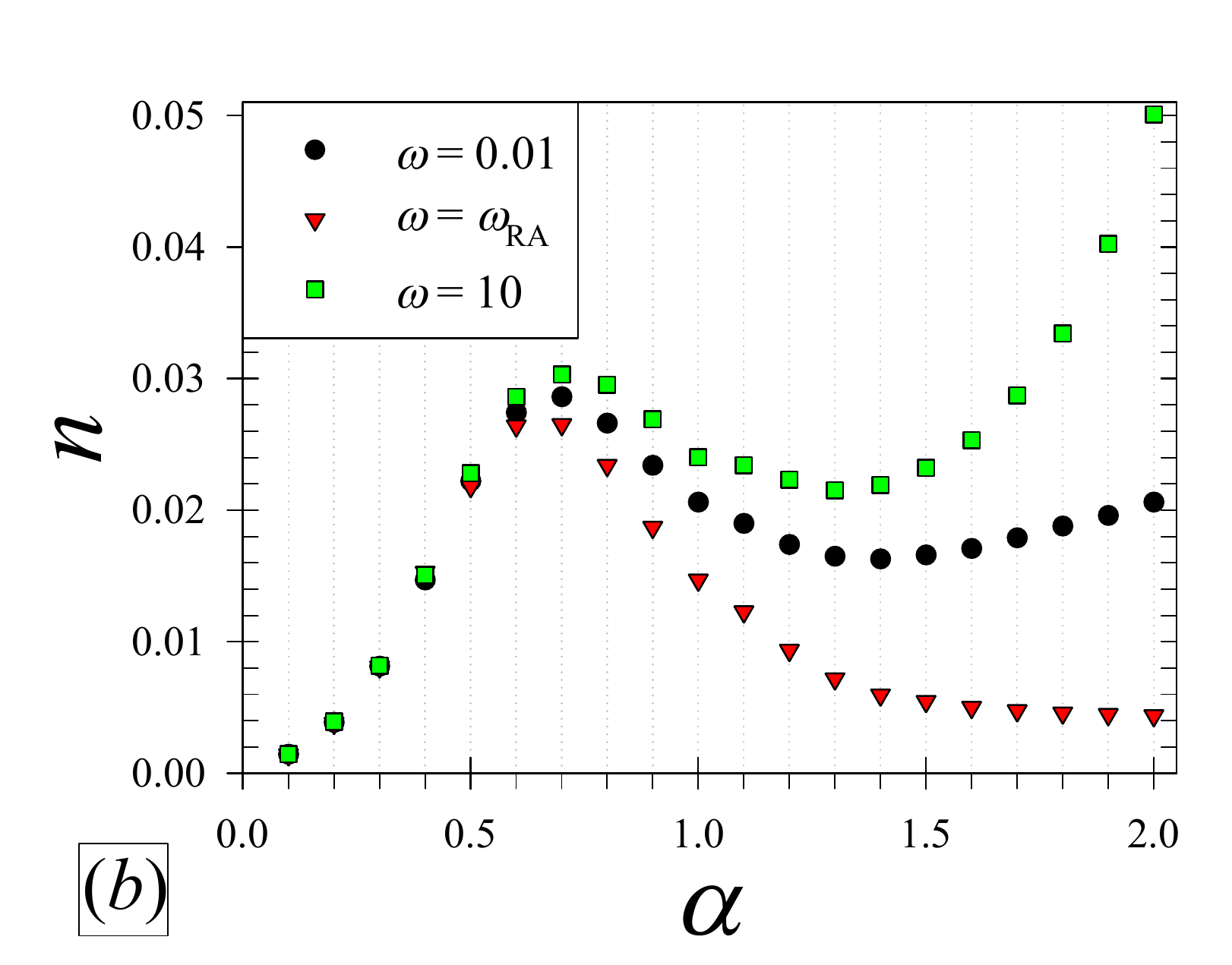}
\caption{(Color online) \emph{a}) MST $\tau$ as a function of the driving frequency $\omega$ for $S_{\alpha}(1, 0, 0)$ and $\alpha\in(0,2]$. The inset shows the values of the MSTs as a function of $\alpha$ for a low frequency $\omega=0.01$, for the frequencies of the RA minima $\omega=\omega_{RA}$, and for a high frequency $\omega=10$. \emph{b}) Mean soliton density $n$ as a function of $\alpha$, for $S_{\alpha}(1, 0, 0)$ and $\alpha\in(0,2]$, and for $\omega=0.01, \omega_{RA},10$. The values of the other parameters, $L=10$, $\gamma=0.1$ and $i_0=0.9$ are shown in panel \emph{a} and refer to both panels.}
\label{Fig5}
\end{figure}
\begin{figure}[t!!]
\centering
\includegraphics[width=0.49\textwidth]{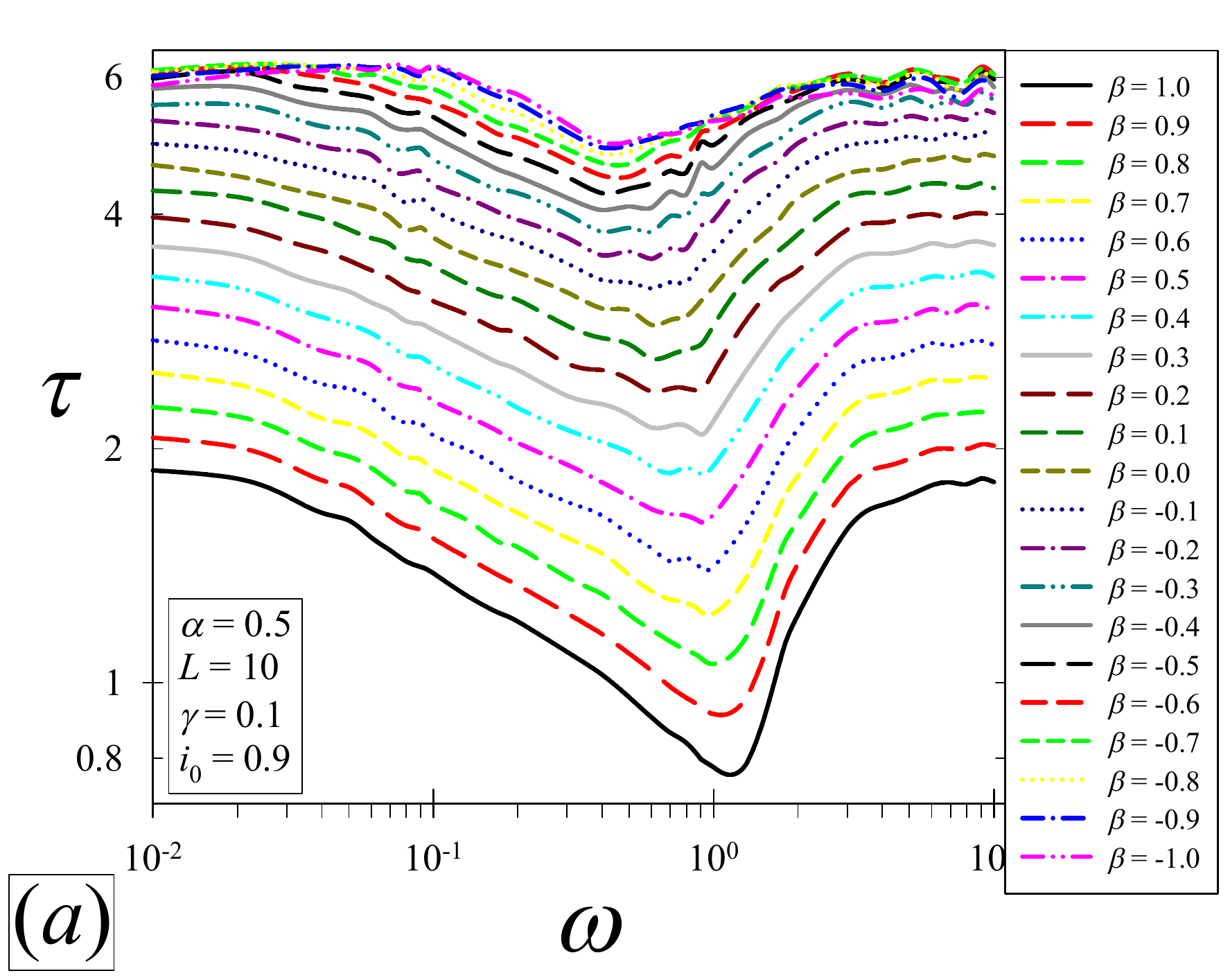}
\includegraphics[width=0.50\textwidth]{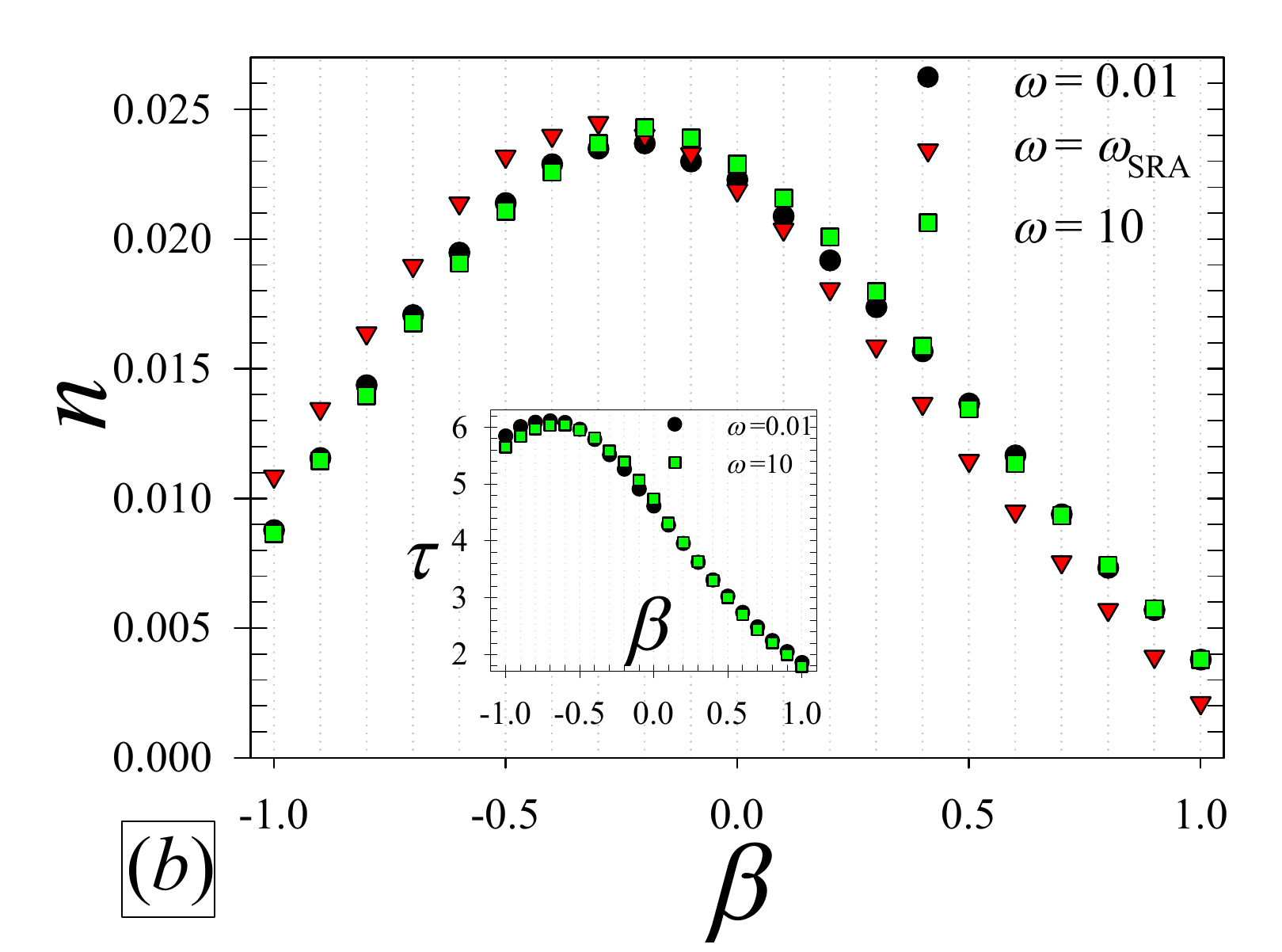}
\caption{(Color online) \emph{a}) MST $\tau$ as a function of the driving frequency $\omega$ for $S_{0.5}(1, \beta, 0)$ and $\beta\in[-1,1]$. \emph{b}) Mean soliton density $n$ as a function of $\beta$ for $S_{0.5}(1, \beta, 0)$ and $\beta\in[-1,1]$, and for $\omega=0.01, \omega_{RA},10$. The inset shows the values of the MSTs as a function of $\beta$ for low and high frequencies, $\omega=0.01$ and $\omega=10$. The values of the other parameters, $L=10$, $\gamma=0.1$, and $i_0=0.9$ are shown in panel \emph{a} and refer to both panels.}
\label{Fig6}
\end{figure}

In this section we analyze the behavior of the MSTs as a function of the driving frequency $\omega$. The results for different values of $\alpha\in(0,2]$ are shown in panel \emph{a} of Fig.~\ref{Fig5} setting $\beta=0.0$, $L=10$, $\gamma=0.1$, and $i_0=0.9$. The junction is long enough for solitons to be observed. \\
\indent All curves of Fig.~\ref{Fig5}\emph{a} clearly show the presence of \emph{resonant activation} (RA)~\cite{Doe92,Man00,Dub04,Man98,Pec94,Mar96,Dyb09,Miy10,Has11,Fia11,Val14}, specifically \textit{stochastic resonance activation}, a noise induced phenomenon, whose signature is the appearance of a minimum in the curve of MST \emph{vs} $\omega$.
When the noise intensities are greater than $\overline{\Delta U}_{i_0}$ (see Eq.~(\ref{DeltaU})), that is the time average over a driving period of the potential barrier height, the minimum tends to vanish reducing $\alpha$ (see Fig.6\emph{c} of Ref.~\cite{Val14}). Accordingly, using $i_0=0.9$ the time average of the potential barrier height is $\overline{\Delta U}_{i_0=0.9} \simeq 0.4$, and we set $\gamma=0.1$. \\
\indent The RA phenomenon is robust enough to be observed also in the presence of L\'{e}vy noise sources~\cite{Aug10,Gua13,Val14,Spa15}.
Particle escape from a potential well is driven when the potential barrier oscillates on a time-scale characteristic of the particle
escape itself. Since the resonant frequency is close to the inverse of the average escape time at the minimum, which is
the mean escape time over the potential barrier in the lowest configuration, \textit{stochastic resonant
activation} occurs~\cite{Add12,Pan09,Val14,Gua15}. This phenomenon is different from the \textit{dynamic resonant activation}, which is observed when the driving frequency matches the natural frequency of the system, i.e. the plasma frequency~\cite{Dev84,Dev85,Mar87,Gua15}. \\
\indent Considering the curves in Fig.~\ref{Fig5}\emph{a}, we note
that the RA minima shift towards higher frequencies
reducing $\alpha$. Moreover, all these curves are characterized by a
non-monotonic behavior as a function of $\alpha$, for frequencies
within the range (0.04, 1.3). This behavior is highlighted in the
inset of Fig.~\ref{Fig5}\emph{a}, that shows the values of the MSTs
as a function of $\alpha$ for low and high frequencies,
$\omega=0.01$ and $\omega=10$, and for the frequency in the RA minima $\omega=\omega_{RA}$. The curve for
$\omega=\omega_{RA}$ has a maximum for $\alpha=0.7$. A similar
behavior is observed looking the mean soliton densities $n$ versus
$\alpha$ presented in panel \emph{b} of Fig.~\ref{Fig5}. This graph
shows results for $\omega=0.01, \omega_{RA},10$. All these
curves have a maximum located in $\alpha_{max}=0.7$. We observe that
changes in the value of the driving frequency slightly affect both
the position and the height of the $n$ maxima, as well as the values
of $n$ for $\alpha\le\alpha_{max}$. Conversely, for
$\alpha>\alpha_{max}$, the curves in Fig.~\ref{Fig5}\emph{b} have
not the same behavior. This can be explained observing that, for
$\alpha\le\alpha_{max}$, the behavior of the string is strongly
related to the heavy tails characteristics of the L\'evy noise, and
the density of solitons rises up with $\alpha$ reaching a maximum
when the probability $P^{\alpha,\gamma}$ is the highest. Further
increasing $\alpha$, the effect of the L\'evy jumps reduces, but
solitons can be still created as a result of escape events after
oscillations of the cells within the minimum. This explains the
frequency-dependence of the mean soliton density $n$ for
$\alpha>\alpha_{max}$ (see Fig.~\ref{Fig5}\emph{b}). For
$\omega=\omega_{RA}$, the resonant escape events are so rapid that
the soliton formation is hindered, and the mean soliton density
tends to low values. For off-resonance frequencies, temporary trapping of the phase occurs. As a consequence, the mean soliton density $n$ increases, and this effect is greater for high frequencies. \\
\indent The MSTs versus $\omega$ for different values of $\beta\in[-1,1]$ are shown in panel \emph{a} of Fig.~\ref{Fig6} setting $\alpha=0.5$, $L=10$, $\gamma=0.1$, and $i_0=0.9$. The RA phenomenon is still present for all values of $\beta$, but the frequency in correspondence of the minimum grows increasing $\beta$.\\
\indent We note that a general lowering of $\tau$ occurs with increasing $\beta$. This behavior is related to the asymmetric fluctuations that these noise sources induce for $\beta\neq 0$. For $\beta<0$, the L\'evy jumps push the string in the negative $\varphi$ direction, that is in the opposite direction with respect to the tilting imposed by the positive bias current. Therefore, the confinement of the string within the initial metastable state is the longest for $\beta=-1$. Conversely, a positive value of $\beta$ gives fluctuations supporting the slipping of the string along the potential, resulting in very low values of $\tau$. \\
\indent Panel \emph{b} of Fig.~\ref{Fig6} shows the mean soliton densities $n$ as a function of $\beta$ for $\omega=0.01, \omega_{RA},10$. These curves are slightly affected by the value of the driving frequency, but have a maximum for $\beta_{max}=-0.2$, that is in correspondence of a L\'evy distribution only slightly skewed to the left. The nonmonotonic behavior of $n$ versus $\beta$ is due to asymmetry properties of the L\'evy noise for $\beta\neq0$. For high skewed distributions, that is $\beta\gtrsim-1$ and $\beta\lesssim1$, the escape processes occur preferentially towards left or right, respectively, along the washboard potential. This gives rise to low values of the mean soliton density $n$. For more symmetrical distributions, that is $\beta\sim0$, the escape events decrease and, as a consequence, $n$ increases. This behavior is related to the nonmonotonic behavior of the MST as a function of $\beta$ at low and high frequencies shown in the inset of Fig.~\ref{Fig6}\emph{b}. At $\omega\approx \omega_{RA}$, the resonant escape process drives the transient dynamics. Moreover, by varying the asymmetry parameter $\beta$ from -1 to 1 the string is pushed towards the same direction of the tilted potential, enhancing the escape process. This gives rise to a monotonic behavior of $\tau$ as a function of $\beta$ at the resonance, see Fig.~\ref{Fig6}\emph{a}.

\subsection{Results as a function of $\gamma$}

\begin{figure}[t]
\centering
\includegraphics[width=\textwidth]{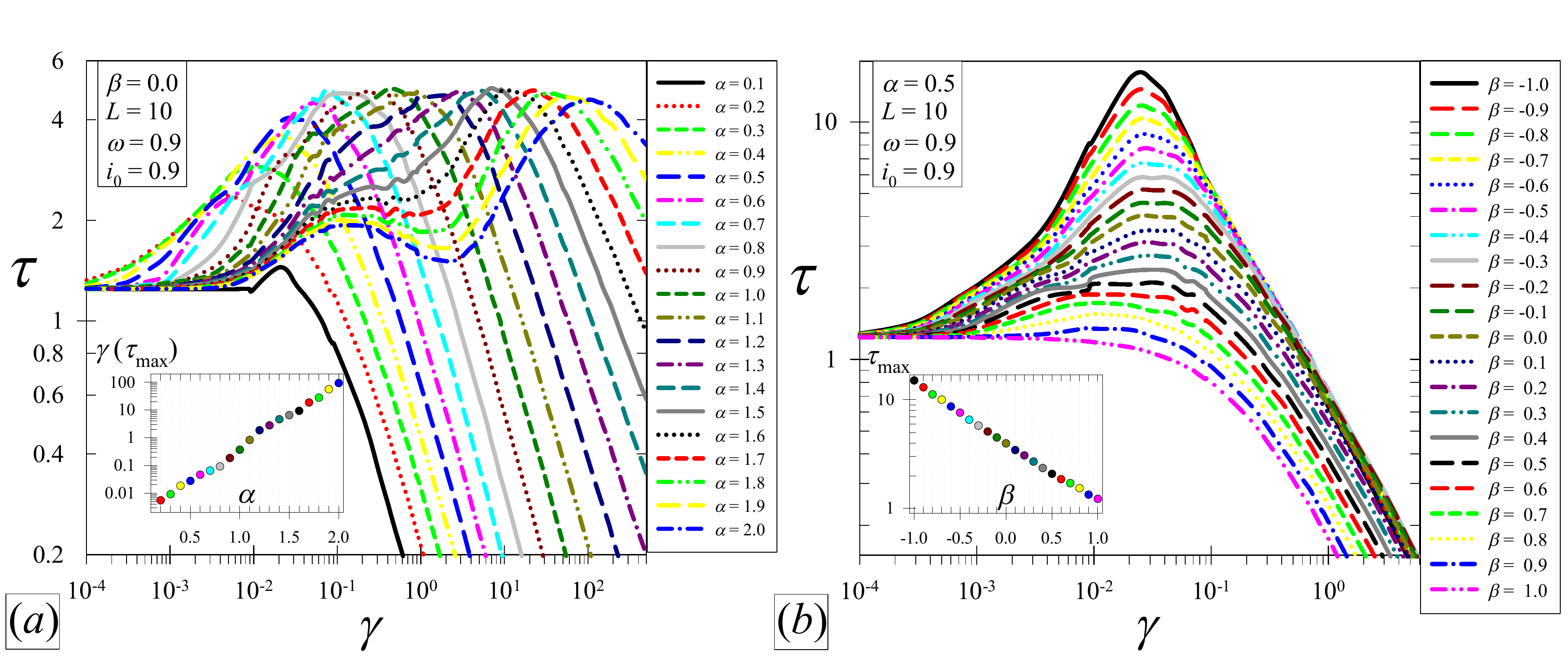}
\caption{(Color online) \textit{a}) MST $\tau$ as a function of the noise amplitude $\gamma$ for $S_{\alpha}(1, 0, 0)$ and $\alpha\in(0,2]$. The inset shows the position $\gamma(\tau_{max})$ of the NES maxima as a function of $\alpha$. \textit{b}) MST $\tau$ as a function of the noise amplitude $\gamma$ for $S_{0.5}(1, \beta, 0)$ and $\beta\in[-1,1]$. The inset shows the height $\tau_{max}$ of the NES maxima as a function of $\beta$. The values of the other parameters are $L=10$, $\omega=0.9$, and $i_0=0.9$.}
\label{Fig7}
\end{figure}

\begin{figure}[t!!]
\centering
\hspace{0.01\textwidth}
\includegraphics[width=0.70\textwidth]{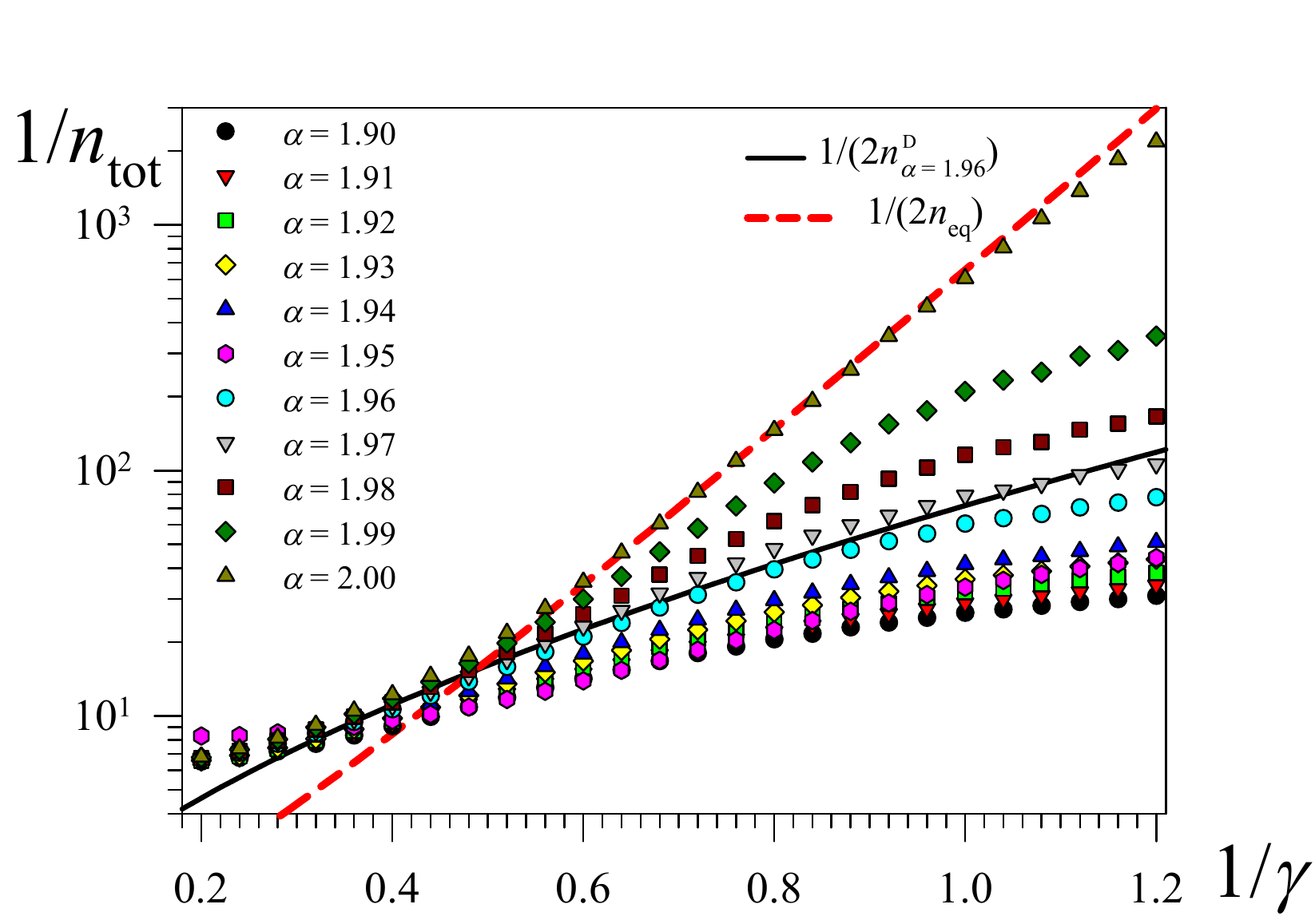}
\caption{(Color online)
Inverse mean density of solitons $1/n_{tot}$ as a function of inverse noise intensity $1/\gamma$ in the absence of bias force $i_b=0$, for $S_{\alpha}(1, 0, 0)$ and $\alpha\in[1.9,2.0]$, and for $L=40$. The red dashed line is the inverse of equilibrium soliton density $2n_{eq}$ calculated according to Eq.~(\ref{Kink_density})~\cite{But79,But81,But95,Fed09} and the black solid line is obtained from Eq.~(\ref{Kink_density}) by replacing $\gamma$ with the L\'evy noise amplitude $D=(2\gamma)^{1/\alpha}$ with $\alpha=1.96$. }
\label{Fig8}
\end{figure}

In this section we analyze the MST versus the non-Gaussian noise amplitude $\gamma\in[10^{-4},5\cdot10^2]$, for different values of $\alpha$. The results for $\alpha\in(0,2]$, $\beta=0.0$, $L=10$, $\omega=0.9$, and $i_0=0.9$ are shown in panel \emph{a} of Fig.~\ref{Fig7}. \\
\indent Regardless the value of $\alpha$, for $\gamma\rightarrow 0$ all the curves converge to the same value, i.e. the deterministic lifetime in the superconducting state, which strongly depends on the bias current.
Increasing the intensity of noise, the MST curves exhibit an effect of \emph{noise enhanced stability} (NES)~\cite{Dub04,Man96,Agu01,Spa04,DOd05,Fia05,Hur06,Spa07,Man08,Yos08,Fia09,Tra09,Fia10,Li10,Smi10,Val14}. This is a noise induced phenomenon consisting in a nonmonotonic behaviour as a function of the noise intensity with the appearance of a maximum. This implies that the stability of metastable states can be enhanced and the average lifetime of the metastable state increases nonmonotonically with the noise intensity.
The observed nonmonotonic resonance-like behavior disagrees with the monotonic behavior of the Kramers theory and its extensions~\cite{Kra40,Mel91,Han90}. This enhancement of stability, first noted by Hirsch et al.~\cite{Hir82}, has been observed in different physical and biological systems, and belongs to a highly topical interdisciplinary research field, ranging from condensed matter physics to molecular biology and to cancer growth dynamics~\cite{Spa07,Spa12}.\\
\indent We observe that in the curve for $\alpha=2$ two maxima are present in $\gamma\simeq0.1$ and $\gamma\simeq100$. Reducing the value of $\alpha$ the higher maximum shifts towards smaller $\gamma$ maintaining its height, up to merge with the first maximum. In particular, the position $\gamma(\tau_{max})$ of the second NES peak grows exponentially towards higher noise intensities as the value of $\alpha$ increases (see inset of Fig.~\ref{Fig7}\emph{a}).\\
\indent The double maxima NES for short and long JJs in the presence of a Gaussian noise source, i.e. $\alpha=2$ and $\beta=0.0$, was previously observed by Valenti \emph{et al.}~\cite{Val14}. In view of understanding the physical motivations of these NES effect, they calculate the time evolution of the probability $\Pt$, as defined in Eq.~(\ref{P_averaged}), during the switching dynamics of the junction (cf. panel \emph{b} of Fig.8 in Ref.~\cite{Val14}). In correspondence of the first maximum, the contemporaneous presence of the oscillating potential and the noise source hinders the phase switching and therefore the passage of the junction to the resistive regime.
The exit from the first well is not sharp and $\Pt$ assumes an oscillatory behavior, almost in resonance with the periodical motion of the washboard potential. This oscillating behavior of $\Pt$ tends to disappear as the noise intensity increases. In correspondence of the second NES peak for higher noise intensities no oscillations in $\Pt$ are present. The JJ dynamics is totally driven by the noise, and the NES effect is due to the possibility that the phase string comes back into the first valley after a first escape event, as indicated by the fat tail of $\Pt$ (cf. Fig.8\emph{b} of Ref.~\cite{Val14}). \\
\indent Reducing $\alpha$, the probability to obtain intense noise fluctuations grows, so that temporary confinements within the initial metastable state after a first escape can still occur but in correspondence of lower intensities of noise. The second NES maximum is therefore present also for $\alpha<2$, as shown in Fig.~\ref{Fig7}\emph{a}.\\
\indent Panel \emph{b} of Fig.~\ref{Fig7} shows the MSTs as a function of $\gamma$, for different values of $\beta\in[-1,1]$, for $\alpha=0.5$, $L=10$, $\omega=0.9$, and $i_0=0.9$. The NES effect is still evident, although it is strongly affected by the $\beta$ value. In fact, the time of confinement of the string in the initial metastable state is longer for a L\'evy noise strongly skewed in the opposite direction with respect to the positive $\varphi$ direction, that is with respect to the tilting imposed by the positive bias current. We observe that only the height of the NES peak if affected by the $\beta$ value, whereas its position is unchanged varying $\beta$. In particular, the inset of Fig.~\ref{Fig7}\emph{b} shows that the height of the NES peak exponentially decreases by increasing $\beta$.\\
\indent Both panels of Fig.~\ref{Fig7} show that, for high noise intensities, the MSTs have a power-law dependence on the noise intensity according to the expression $\tau \simeq C(\alpha)/\gamma^{\mu(\alpha)}$, where both the prefactor C and the exponent $\mu$ depend on the L\'evy index $\alpha$~\cite{Dub08}.\\
\indent From Fig.~\ref{Fig7}, we obtain $\mu(\alpha) \in (0.7,1.1)$ for $\alpha$ in $(0,2]$ and $\beta=0.0$ (see curves in panel \emph{a}) and $\mu(\alpha) \sim 1.1$ for $\alpha=0.5$ and $\beta\in[-1,1]$ (see curves in panel \emph{b}). These values of $\mu(\alpha)$ are in agreement with the exponent $\mu(\alpha) \approx 1$ for $0 < \alpha < 2$, calculated for barrier crossing in bistable and metastable potential profiles~\cite{Che05,Che07}.\\
\indent Moreover, we calculate the total mean soliton density for the zero bias force, that is $i_0=i_{ac}=0$. In this condition, the solitons are exclusively generated by noise induced fluctuations along the string. The observation time is increased to $t_{max}=10^3$ and the junction length is $L=40$. The results are presented in Fig.~\ref{Fig8}, where the inverse of the mean soliton density $1/n_{tot}$ is plotted as a function of the inverse noise intensities $1/\gamma$ for $\alpha\in[1.9,2.0]$ and $\beta=0.0$. We can compare the curve for $\alpha=2$ with the formula for the kink density~\cite{But79,But81,But95,Fed09}
\begin{equation}
n_{eq}=\sqrt{\frac{2}{\pi}\frac{E_k}{\gamma}}\exp\left ( -\frac{E_k}{\gamma} \right ).
 \label{Kink_density}
\end{equation}
\indent The total soliton density (kinks and antikinks) in a string is twice the kink density given in Eq.~(\ref{Kink_density}) (see $1/\left ( 2n_{eq} \right )$ curve, red dashed line, in Fig.~\ref{Fig8}). In Eq.~(\ref{Kink_density}), the energy $E_k=8E_J$ is the rest energy of a kink (or an antikink). The total soliton density $n_{tot}$ calculated for $\alpha=2$ perfectly matches the prediction of Eq.~(\ref{Kink_density}) for $1/\gamma\gtrsim0.5$. \\
\indent When the noise intensity is greater than the mean potential barrier height $\overline{\Delta U}_{i_0=0.0}=2$, i.e. $1/\gamma \lesssim0.5$, the curves in Fig.~\ref{Fig8} overlap.
For smaller noise intensities, i.e. $1/\gamma>0.5$, the mean soliton density $n_{tot}$ definitively increases as the value of $\alpha$ slightly reduces. We observe that the behavior of the curves for $\alpha\in[1.9,2.0)$ can be approximately estimated by replacing in Eq.~(\ref{Kink_density}) the noise intensity with the L\'evy noise amplitude $D^{\alpha,\gamma}=(2\gamma)^{1/\alpha}$ (see black solid line in Fig.~\ref{Fig8} for $\alpha=1.96$).

\section{Conclusions}

\indent We studied how both stochastic non-Gaussian fluctuations and an oscillating driving contribute to the generation of solitons in a long Josephson junction (also called string), which is a system governed by the sine-Gordon model. The non-Gaussian noise sources are modeled by using $\alpha$-stable L\'evy statistics. In detail, we analyzed the behavior of the superconducting lifetime and the voltage drop across the junction as the values of the characteristic L\'evy parameters $\alpha$ and $\beta$ change. The superconducting lifetime of the junction is calculated as the mean switching time (MST) from the initial metastable state, i.e. a minimum of the tilted washboard potential.\\
\indent Studying the MST as a function of the junction length $L$, two different behaviors are observed, in correspondence of regimes of length below and above a critical value $L_c$. One, occurring for short junctions, is characterized by the movement of the phase string as a whole. The other one, occurring for junctions whose size exceeds $L_c$, in which the solitons creation is allowed. We found a connection between the behavior of the MST for $L>L_c$ and the mean density $n$ of solitons partially laying in the initial metastable state. Moreover, we observed that, for a fixed length of the junction, $n$ varies non-monotonically as a function of $\alpha$, showing a clear maximum. A similar non-monotonic behavior characterizes the probability $P^{\alpha,\gamma}$ of a direct ``noise-induced'' jump from the initial washboard minimum to the next one. \\
\indent We investigated also the mean potential difference (MPD) as
a function of the junction length. For $L>L_c$ the MPD behaves just
like the mean density $n_{tot}$ of solitons formed along the string.\\
\indent Studying the behavior of the MST as a function of the
driving frequency $\omega$, we observed the presence of stochastic
resonance activation, that is a noise induced phenomenon whose
signature is the appearance of a minimum in the curve of MST versus
$\omega$. For frequencies nearby the RA minima, the behavior of the
MST as a function of $\alpha$ is characterized by the presence of a
maximum. We understood this result studying the jump probability
$P^{\alpha,\gamma}$, which shows the same nonmonotonic behavior as a
function of $\alpha$.\\
\indent We found evidence of a nonmonotonic behavior also studying
the MSTs as a function of the noise intensity $\gamma$, observing
the phenomenon of noise enhanced stability, whose characteristics
strongly depend on the values of the L\'evy parameters $\alpha$ and
$\beta$.\\
\indent Our findings are important to understand the role of non Gaussian noise sources on the transient dynamics of out of equilibrium long JJ devices. 
This is important not only in the general context of the nonequilibrium statistical mechanics, but also to improve the performance of this device. 
Moreover, the statistics of the escape times from the superconductive metastable state of a JJ carries information on the non-Gaussian background
noise~\cite{Val14}.


\providecommand{\newblock}{}

\end{document}